\documentclass[aps,prd,preprint,tightenlines,superscriptaddress]{revtex4}
\usepackage{epsfig}
\usepackage{graphicx}
\usepackage{psfrag}
\usepackage{amsmath,amssymb}
\usepackage{colordvi}
\usepackage{color}
\usepackage{amsfonts}
\usepackage{enumerate}
\usepackage{slashed,bm}

\def \non {\nonumber}
\def \beq  {\begin{equation}}
\def \eeq  {\end{equation}}

\begin{document}

\title{Parton distribution functions from reduced Ioffe-time distributions}
\author{Jian-Hui Zhang}
\affiliation{Institut f\"ur Theoretische Physik, Universit\"at Regensburg, \\
D-93040 Regensburg, Germany}
\author{Jiunn-Wei Chen}
\affiliation{Department of Physics, Center for Theoretical Physics, and Leung Center for Cosmology and Particle Astrophysics, National Taiwan University, Taipei, Taiwan 106}
\author{Christopher Monahan}
\affiliation{Institute for Nuclear Theory, University of Washington, Seattle, Washington 98195, USA}
\vspace{0.5in}
\begin{abstract}

We show that the correct way to extract parton distribution functions from the reduced Ioffe-time distribution (RITD), a ratio of the Ioffe-time distribution for a moving hadron and a hadron at rest, is through a factorization formula. This factorization exists because, at small distances, forming the ratio does not change the infrared behavior of the numerator, which is factorizable. We illustrate the effect of such a factorization by applying it to results in the literature.
\end{abstract}

\preprint{INT-PUB-17-057}

\maketitle

\section{Introduction}
One of the most important goals of quantum chromodynamics (QCD) is to understand the structure of hadrons in terms of their fundamental constituents -- quarks and gluons. This is a profoundly difficult task because it necessarily requires studies in the non-perturbative regime. Currently, the most reliable tool for such non-perturbative studies is lattice QCD, which has been widely used to study the static property of hadrons (see e.g.~\cite{Edwards:2012fx,Aoki:2016frl}). However, lattice QCD is formulated in Euclidean spacetime and so can not be used to calculate quantities with real-time dependence. For example, parton distribution functions (PDFs), which characterize the momentum distributions of quarks and gluons inside the hadron, are defined in terms of lightcone correlations and therefore have real-time dependence. Traditionally, the PDFs can only be probed indirectly on the lattice by calculating their moments~\cite{Detmold:2001jb,Dolgov:2002zm}. Beyond the first few moments such calculations become very hard due to technical difficulties associated with the symmetry of the lattice regulator, which leads to power-divergent mixing between different moments.

Recently, a new approach has been proposed to study parton physics from lattice QCD~\cite{Ji:2013dva}, which allows a direct computation of the full-$x$ dependence of parton quantities such as the PDFs. This approach has been formulated as the large momentum effective theory (LaMET)~\cite{Ji:2014gla}. The LaMET is based on the observation that parton physics defined in terms of lightcone correlations can be obtained from time-independent spatial correlations, now known as quasi-distributions, boosted to the infinite momentum frame. The quasi-distributions can be calculated on the lattice, since they are time-independent quantities. For finite but large momenta feasible on the lattice, the Euclidean quasi-distributions can be related to the physical ones by a factorization formula.  The LaMET has been applied to computing  the nucleon unpolarized, helicity and transversity PDFs~\cite{Lin:2014zya,Alexandrou:2015rja,Alexandrou:2016jqi,Chen:2016utp,Chen:2017mzz,Lin:2017ani,Alexandrou:2017dzj}, as well as the meson distribution amplitudes (DAs)~\cite{Zhang:2017bzy,Chen:2017gck} (for related studies see Refs.~\cite{Xiong:2013bka,Ji:2014hxa,Xiong:2017jtn,Wang:2017qyg,Stewart:2017tvs,Ji:2015qla,Xiong:2015nua,Radyushkin:2017ffo,
Ji:2015jwa,Ishikawa:2016znu,Chen:2016fxx,Constantinou:2017sej,Alexandrou:2017huk,Ji:2017rah,Ji:2017oey,Ishikawa:2017faj,
Green:2017xeu,Li:2016amo,Jia:2015pxx,Gamberg:2014zwa,Monahan:2016bvm,Carlson:2017gpk,Briceno:2017cpo,
Rossi:2017muf,Monahan:2017oof,Broniowski:2017wbr,Broniowski:2017gfp,Chen:2017mie,Chen:2017lnm,Wang:2017eel}), and led to fairly encouraging results.

Alternative, but related, approaches have also been proposed: using lattice cross sections \cite{Ma:2014jla,Ma:2017pxb} and the Ioffe-time distribution or pseudo-distribution \cite{Radyushkin:2017cyf} to extract the $x$-dependence of PDFs from lattice QCD. In particular, the pseudo-distribution essentially corresponds to a different Fourier transform of the same equal-time Euclidean correlation function as used in LaMET. The pseudo-distribution proposal is therefore equivalent to the quasi-distribution calculation in LaMET, in the sense that similar factorization formulae exist for both approaches and that large nucleon momentum is required in both cases to extract the information on the leading-twist PDF~\cite{Ji:2017rah}. In Ref.~\cite{Radyushkin:2017cyf} it was claimed that, unlike the quasi-distribution, the pseudo-distribution only has a support in the physical $x$ range ($x=[-1,1]$) as a result of Fourier transform over Ioffe-time. However, that relies on the integration over an infinite range of Ioffe-time. In practical lattice simulations, the Ioffe-time that can be reached is always limited to a finite range. Therefore, the pseudo-distribution may have residual contributions in unphysical region as well, similar to the quasi-distribution. In addition to the above methods, there have been a number of proposals to use current-current correlators at spacelike separation to compute PDFs, the pion DA, and related quantities~\cite{Davoudi:2012ya,Detmold:2005gg,Liu:1993cv,Liang:2017mye,Braun:2007wv,Bali:2017gfr,Chambers:2017dov}.

In Ref.~\cite{Orginos:2017kos,Karpie:2017bzm}, an exploratory study of extracting quark PDFs from the Ioffe-time distribution/pseudo-distribution was carried out. The authors used a reduced Ioffe-time distribution (RITD), formed by taking the ratio of the equal-time correlator for a moving hadron and a hadron at rest. The advantage of the RITD is that the renormalization factors associated with the Ioffe-time distribution~\cite{Braun:1994jq} cancel between the numerator and the denominator, since they have been shown to be independent of the nucleon momentum~\cite{Ji:2017oey,Ishikawa:2017faj,
Green:2017xeu}. Therefore, the uncertainties related to non-perturbative renormalization effects, such as those seen in Ref.~\cite{Chen:2017mzz}, for example, can be avoided. The authors of Ref.~\cite{Orginos:2017kos} then determined the PDFs based on an approximate scaling in $z^2$ observed in the lattice data of the RITD, where $z$ is the distance between the two quark fields defining the equal-time quark correlator. They also highlighted the remaining $z^2$-dependence of the RITD and studied its evolution with $z^2$ (see also~\cite{Radyushkin:2017lvu}).

In this paper, we show that the correct way to take advantage of the RITD is through a factorization formula that connects the RITD to the PDFs. Although logarithmic evolution relates the RITD at different $z^2$, a factorization is required to convert the RITD to the PDFs. The reason that such a factorization exists is, as we will show later, that the RITD has the same infrared (IR) behavior as its numerator, which is factorizable, provided that $z^2$ is small.

The paper is organized as follows. In Sec.~\ref{facRITD}, we present the factorization for the RITD, and give the hard kernel connecting the RITD and the PDFs at one-loop order. As a demonstration, we then apply the factorization to the data read from the figures in Ref.~\cite{Orginos:2017kos} in Sec.~\ref{appfac}. A summary is given in Sec.~\ref{concl}.

\section{Factorization of Reduced Ioffe-time distribution}
\label{facRITD}
The RITD introduced in Ref.~\cite{Radyushkin:2017cyf} is defined as
\beq\label{RITDdef}
M(\nu=z\cdot p, z^2, \mu)=\frac{\mathcal M(\nu, z^2, \mu)}{\mathcal M(0, z^2, \mu)},
\eeq
where 
\beq
\mathcal M(\nu, z^2, \mu)=\langle p|\bar\psi(z)\Gamma L(z,0)\psi(0)|p\rangle
\eeq
with $\Gamma$ a Dirac matrix and $L(z,0)$ the straight line gauge link connecting the two quark fields. $\mu$ denotes the renormalization scale. The matrix element $\mathcal M(\nu, z^2, \mu)$ is the same as used in LaMET to calculate quasi-distributions, and can be interpreted as a covariant Ioffe-time distribution with a given Ioffe-time $\nu=-z\cdot p$ at a spacelike separation $z^2<0$. For the unpolarized quark distribution, in order that the denominator of Eq.~(\ref{RITDdef}) does not vanish at leading power, $\Gamma$ can chosen as $\Gamma=\gamma^0$.

The quasi- and the pseudo-distribution are defined as two different Fourier transforms of the Ioffe-time distribution $\mathcal M(\nu, z^2, \mu)$, the former with respect to $z$ and the latter with respect to $\nu$. The authors of Ref.~\cite{Ji:2017rah} demonstrated that there exists a similar factorization formula for the pseudo-distribution as for the quasi-distribution, but at small distance instead of large momentum. For a more explicit demonstration of the equivalence between the small distance factorization in the pseudo-distribution and the large momentum factorization in the quasi-distribution, see~\cite{Zhao}. In fact, both factorizations originate from the same factorization of Ioffe-time distribution, which can be easily obtained from the momentum space factorization of the pseudo-distribution \cite{Ji:2017rah}
\begin{equation}
\mathcal P(y, z^2, \mu) = \int \frac{dx}{|x|}Z\Big(\frac{y}{x},z^2\mu^2\Big)q(x,\mu)+\mathcal O(z^2).
\end{equation}
Here
\beq
\mathcal P(y, z^2, \mu)=\frac{1}{2\pi}\int d\nu\, e^{-iy\nu} \mathcal M(\nu, z^2, \mu)
\eeq
is the pseudo-distribution, $q(x,\mu)$ is the PDF, $\mu$ denotes a renormalization scale in the $\overline{MS}$ scheme, and $Z$ is the perturbative hard matching kernel. For simplicity, we have chosen the same renormalization scale in $\mathcal P(y, z^2, \mu)$ and $q(x, \mu)$. Taking the Fourier transform with respect to the Ioffe-time, $\nu$, leads to
\begin{align}\label{coordfac}
\mathcal M(\nu, z^2, \mu)&=\int dy\, e^{iy\nu}\mathcal P(y, z^2, \mu)=\int dy\, e^{iy\nu}\int \frac{dx}{|x|}Z\Big(\frac{y}{x},z^2\mu^2\Big)q(x,\mu)+\mathcal O(z^2)\non\\
&=\int dy\, e^{iy\nu}\int \frac{dx}{|x|}Z\Big(\frac{y}{x},z^2\mu^2\Big)\int \frac{d\nu'}{2\pi} \, e^{-i x\nu'}\mathcal M(\nu',0,\mu)+\mathcal O(z^2)\non\\
&=\int du\, \delta\Big(\frac{y}{x}-u\Big)\int dy\, e^{iy\nu}\int \frac{dx}{|x|}Z\Big(\frac{y}{x},z^2\mu^2\Big)\int \frac{d\nu'}{2\pi} \, e^{-i x\nu'}\mathcal M(\nu',0,\mu)+\mathcal O(z^2)\non\\
&=\int du\,Z(u,z^2\mu^2)\mathcal M(u\nu,0,\mu)+\mathcal O(z^2).
\end{align}
Up to $\mathcal O(\alpha_s)$, we can read off the form of $Z$ from Ref.~\cite{Ji:2017rah} as
\begin{align}\label{1loopmatching}
Z(u, z^2\mu^2)&=\delta(1-u)\Big[1+\frac{\alpha_s C_F}{2\pi}\big(\frac{3}{2}\ln(-z^2\mu^2e^{2\gamma_E}/4)+\frac{3}{2}\big)\Big]\non\\
&\hspace{-3em}+\frac{\alpha_s C_F}{2\pi}\Big[-\big(\frac{1+u^2}{1-u}\big)_+(\ln(-z^2\mu^2e^{2\gamma_E}/4)+1)
-\Big(\frac{4\ln(1-u)}{1-u}\Big)_++2(1-u)\Big]\theta(u)\theta(1-u).
\end{align}

So far we have considered the Ioffe-time distribution only. Now let us turn to the reduced distribution. The matrix element $\mathcal M(\nu, z^2, \mu)$ has been shown to renormalize multiplicatively as~\cite{Ji:2017oey,Ishikawa:2017faj,
Green:2017xeu} (see also~\cite{Dotsenko:1979wb,Dorn:1986dt,Craigie:1980qs,Musch:2010ka,Ishikawa:2016znu,Chen:2016fxx,Alexandrou:2017huk})
\beq
\mathcal M_R(\nu, z^2, \mu)=Z_{\bar j}^{-1} Z_j^{-1}e^{-\delta m |z|}\mathcal M_B(\nu, z^2),
\eeq
where $\delta m$ is an effective mass counterterm removing power divergences in the Wilson line, and $Z_{j}, Z_{\bar j}$ are the renormalization factors associated with the endpoint of the Wilson line and independent of $z, p$. Thus, the entire renormalization is independent of the external momentum $p$. By forming the RITD, the renormalization factors completely cancel between the numerator and the denominator. Therefore, the RITD has the potential to significantly reduce uncertainties related to renormalization effects, such as those seen in Ref.~\cite{Chen:2017mzz}, for example.

From Eq.~(\ref{coordfac}), we have
\beq\label{denomint}
\mathcal M(0, z^2, \mu)=\int du\, Z(u, z^2\mu^2)\mathcal M(0,0)+\mathcal O(z^2)=\int du\, Z(u, z^2\mu^2)+\mathcal O(z^2),
\eeq
since $\mathcal M(0,0)=\int dx\, q(x)=1$, which expresses conservation of quark number. In other words, for a small $z^2$, where higher-twist contributions can be neglected, $\mathcal M(0, z^2, \mu)$ depends on the perturbative matching kernel only, and thus does not introduce any IR divergences. This can be easily checked from the one-loop result in Ref.~\cite{Ji:2017rah}. After renormalization, we have
\beq\label{M0z2}
\mathcal M(0,z^2, \mu)=1+\frac{\alpha_s C_F}{2\pi}\Big[\frac{3}{2}\ln(-z^2\mu^2e^{2\gamma_E}/4)+\frac 5 2\Big]+\mathcal O(z^2)
\eeq 
for a small $z^2$, which can also be obtained from Eqs.~(\ref{1loopmatching}) and (\ref{denomint}). Although the UV divergences have been removed from Eqs.~(\ref{1loopmatching}) and (\ref{M0z2}), in fact one can directly use the unrenormalized results and explicitly show that the UV poles cancel in the RITD. Eq.~(\ref{M0z2}) tells us that at small $z^2$, $\mathcal M(0,z^2, \mu)$ only changes the UV behavior and not the IR behavior.

These results suggest the correct way to relate the PDF and the RITD. That is, we can form a factorization not for the Ioffe-time distribution, but for the reduced distribution. This is possible because the RITD is constructed from a ratio of two matrix elements, $\mathcal M(\nu, z^2, \mu)$ and $\mathcal M(0, z^2, \mu)$, where the latter does not introduces IR divergence, therefore the ratio will have the same IR behavior as $\mathcal M(\nu, z^2, \mu)$, which we know is factorizable (see Eq.~(\ref{coordfac}) for its factorization). The factorization of the RITD can be written as
\begin{align}\label{reducedITmat}
M(\nu, z^2, \mu)&=\int du\, \bar Z(u,z^2\mu^2)M(u\nu,0,\mu)+\mathcal O(z^2)\non\\
&=\int d\nu'\, \delta(u\nu-\nu')\int du\, \bar Z(u,z^2\mu^2)M(u\nu,0,\mu)+\mathcal O(z^2)\non\\
&=\int \frac{d\nu'}{|\nu|}\bar Z\Big(\frac{\nu'}{\nu}, z^2\mu^2\Big)M(\nu',0,\mu)+\mathcal O(z^2),
\end{align}
where on the r.h.s., we can use either $M(\nu', 0,\mu)$ or $\mathcal M(\nu', 0,\mu)$ since they are equal. In contrast to the factorization of $\mathcal M(\nu,z^2, \mu)$, we only need to include the extra perturbative corrections from $\mathcal M(0, z^2, \mu)$ to the matching kernel. Note that the renormalization effects cancel between the numerator and the denominator in the RITD, and we can therefore use the above results in any convenient scheme, e.~g.~the $\overline{MS}$ scheme in the continuum. The one-loop result  $\bar Z$ is then given by
\begin{align}\label{reduced1loopmatching}
\bar Z(u, z^2\mu^2)&=\delta(1-u)+\bar Z^{(1)}(u, z^2\mu^2)=\delta(1-u)\Big[1-\frac{\alpha_s C_F}{2\pi}\Big]+\non\\
&\hspace{-2em}\frac{\alpha_s C_F}{2\pi}\Big[-\big(\frac{1+u^2}{1-u}\big)_+(\ln(-z^2\mu^2e^{2\gamma_E}/4)+1)
-\big(\frac{4\ln(1-u)}{1-u}\big)_++2(1-u)\Big]\theta(u)\theta(1-u),
\end{align}
where the extra term $-\frac{\alpha_s C_F}{2\pi}$ in the first row can also be combined with $\frac{\alpha_s C_F}{2\pi}[2(1-u)]$ in the second row to form a complete plus function.

By taking the Fourier transform of Eq.~(\ref{reducedITmat}) with respect to $\nu$ or $z$, we obtain two momentum space representations. One is the pseudo-distribution, the other is related to the quasi-distribution. In the first case, we have (for simplicity, we ignore the higher-twist contributions of $\mathcal O(z^2)$ below, and a detailed investigation of the structure of higher-twist contributions will be given elsewhere~\cite{Braun})
\begin{align}\label{ppdf}
\frac{\mathcal P(x, z^2, \mu)}{\mathcal M(0,z^2, \mu)}&=\int \frac{d\nu}{2\pi}\, e^{-ix\nu}\int\frac{d\nu'}{|\nu|}\bar Z\Big(\frac{\nu'}{\nu}, z^2\mu^2\Big)\int dy\, e^{iy\nu'}q(y, \mu)\non\\
&=\int \frac{dy}{|y|} \bar Z\Big(\frac x y, z^2\mu^2\Big)q(y, \mu).
\end{align}

In the second case, we have
\begin{align}\label{qpdf}
\int \frac{dz}{2\pi}p^z\, e^{-ix\nu}M(\nu,z^2, \mu)&=\int \frac{dz}{2\pi}p^z\, e^{-ix p^z z}\int dy\, e^{iy p^z z}\tilde q(y,\mu)\int dy'\,e^{iy' p^z z} F(y',\mu)\non\\
&=\int dy\, \tilde q(y,\mu)F(x-y,\mu)\non\\
&=\int dy \tilde Z(x, y,\mu)q(y,\mu)
\end{align}
with
\begin{align}
F(y,\mu)&=\int \frac{dz}{2\pi}p^z\, e^{-i y p^z z} \frac{1}{M(0, z^2, \mu)}, \non\\
\tilde Z(x, y,\mu)&=\int du\int \frac{dz}{2\pi}p^z\, e^{-i(x-uy)p^z z}\bar Z(u,z^2\mu^2).
\end{align}

We can take advantage of the RITD either by working directly in coordinate space, convert it to $M(\nu,0, \mu)$ with Eq.~(\ref{reducedITmat}), and then Fourier transform to momentum space to get the PDF; or use Eq.~(\ref{ppdf}) or (\ref{qpdf}) to extract $q(x,\mu)$, where in the latter case we will need a Fourier transform of the matching kernel and an extra convolution of the quasi-distribution.

\section{Application of the RITD factorization formula}
\label{appfac}

The authors of Ref.~\cite{Orginos:2017kos,Karpie:2017bzm} performed an exploratory lattice study to extract the PDF from the RITD, where the computations were performed in the quenched approximation. The authors observed an approximate scaling behavior in $z^2$ of the lattice data for the RITD, then took the RITD approximately as $M(\nu, 0, \mu)$ (they also highlighted the remaining $z^2$-dependence of the RITD and studied its evolution with $z^2$), and tried to fit the results with a functional form of the PDF. They found that the lattice data of the RITD could be well described by the following valence quark distribution
\beq\label{origresult}
u_v(x)-d_v(x)=\frac{315}{32}\sqrt x(1-x)^3.
\eeq 

In the following, we apply the factorization formula in Eq.~(\ref{reducedITmat}) with the one-loop matching kernel in Eq.~(\ref{reduced1loopmatching}) to the data on the RITD in Ref.~\cite{Orginos:2017kos}. The data points we use are extracted from the figures of Ref.~\cite{Orginos:2017kos}.

\begin{figure}[tbp]
\centering
\includegraphics[width=0.6\textwidth]{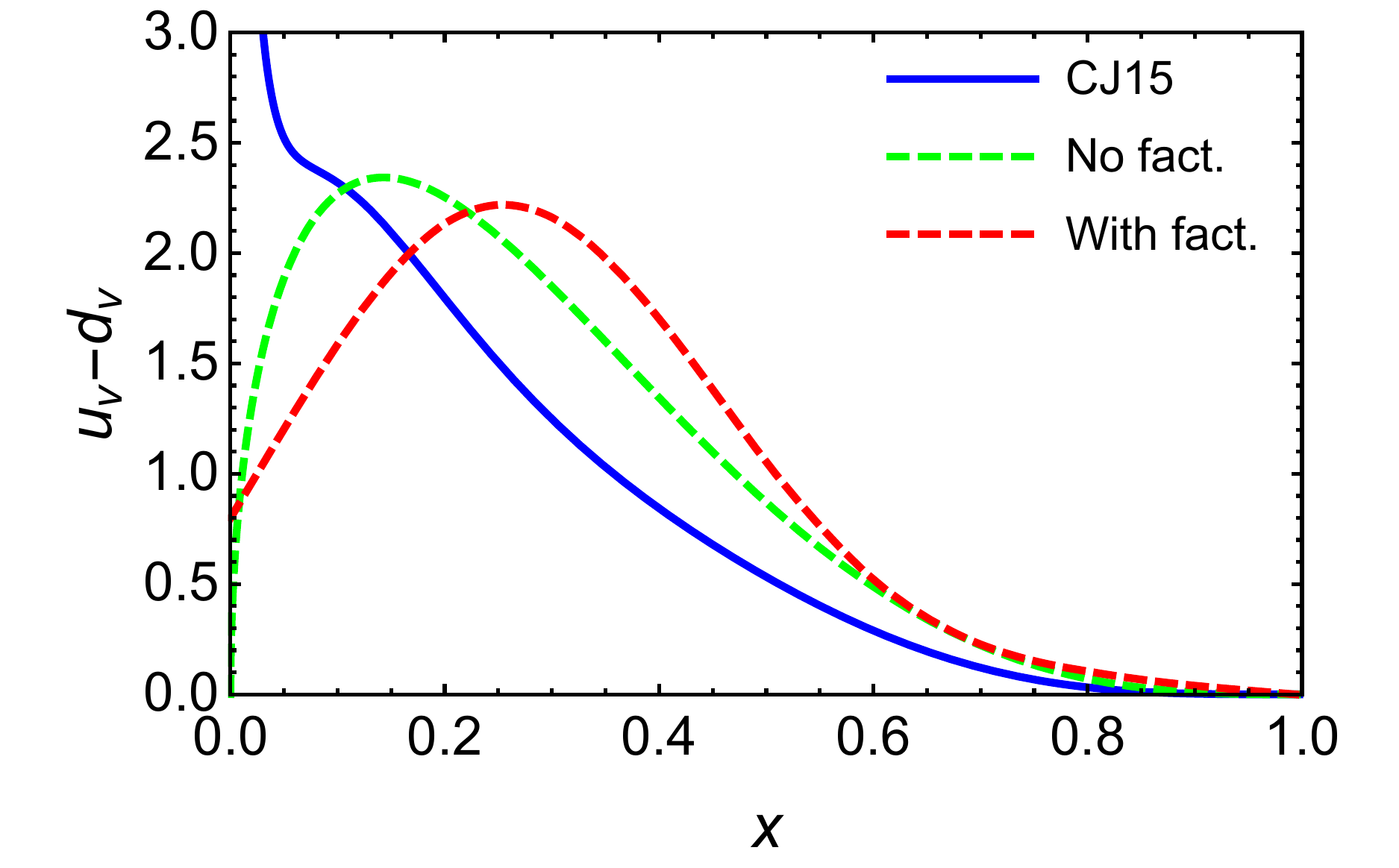}
\caption{Valence distribution $u_v(x)-d_v(x)$ extracted from the data in Ref.~\cite{Orginos:2017kos}. The solid blue curve is the CJ15 global fit~\cite{Accardi:2016qay}, the dashed green curve is the result of Ref.~\cite{Orginos:2017kos} by a fitting to the RITD $M(\nu, z^2, \mu)$, where the authors ignored the factorization converting $M(\nu, z^2, \mu)$ to $M(\nu, 0, \mu)$. The dashed red curve is the result obtained using the factorization formula Eq.~(\ref{reducedITmat}). The data points we use are extracted from the figures of Ref.~\cite{Orginos:2017kos}.}
\label{valencepdf}
\end{figure}

From Eq.~(\ref{reducedITmat}), we have
\beq\label{extraction}
M(\nu, 0, \mu)\approx M(\nu, z^2, \mu)-\int_0^1 du \bar Z^{(1)}(u, z^2\mu^2)M(u \nu, z^2, \mu)
\eeq
up to errors of $\mathcal O(\alpha_s^2)$. The quark distribution can then be obtained by taking the Fourier transform of $M(\nu, 0,\mu)$. As discussed in Ref.~\cite{Orginos:2017kos}, the valence quark distribution is directly related to the real part of $M(\nu, 0,\mu)$. In Fig.~\ref{valencepdf} we show a comparison of the results. The solid blue curve is the CJ15 global fit result~\cite{Accardi:2016qay}. The dashed green curve is the result of Ref.~\cite{Orginos:2017kos} obtained by a numerical fit to the real part of the RITD $M(\nu, z^2,\mu)$, and the factorization converting $M(\nu, z^2,\mu)$ to $M(\nu, 0, \mu)$ in Eq.~(\ref{reducedITmat}) was ignored. The dashed red curve is the result using the factorization formula with $z=4a$ ($a=0.093 fm$), which is the largest $z$ used in Ref.~\cite{Orginos:2017kos} to study the perturbative evolution of the RITD with $z^2$. In principle, it is better to use smaller $z^2$ so that higher-twist contributions are less important. However, for a smaller $z^2$ the Ioffe-time that can be reached also becomes smaller, and therefore will contain less of the information needed to reconstruct the PDF. 

With the choice $z=4a$, the curve generated by applying the one-loop factorization formula, plotted in Fig.~\ref{valencepdf}, lies to the right of the fit results in Ref.~\cite{Orginos:2017kos} and is therefore further away from the phenomenological fit. This is because, at this scale, the term involving $\ln(-z^2\mu^2)$ in the one-loop matching kernel is negative for $z=4a$, the effect of one-loop matching is to push the quark distribution towards the right. Since this logarithm depends on the scale, at finer lattice spacings and larger nucleon momentum, we anticipate that one can use data with smaller $z$, but still with a reasonable range of Ioffe-time, and the matching kernel will then move the result towards the phenomenological curve. We note that this preliminary study is in the quenched approximation, and that we use a matching kernel correct to one-loop in perturbation theory at a relatively low scale, and that therefore quantitative comparison to PDFs extracted from global fits is still not possible with precision. 

\section{Conclusion}
\label{concl}
The reduced Ioffe-time distribution (RITD) has the advantage that the non-perturbative renormalization effects associated with the Ioffe-time distribution itself can be avoided. In this paper, we have shown that the light-front PDFs are related to the RITD through a factorization relation. This factorization exists because, at small distances, forming the ratio in the RITD does not change the IR behavior of the numerator, which is factorizable.  We then illustrate the effect of such a factorization by applying it to the data used in Ref.~\cite{Orginos:2017kos}.

\vspace{1em}
This work was partially supported by the U.S. Department of Energy via grant DE-FG02-00ER41132, 
the SFB/TRR-55 grant ``Hadron Physics from Lattice QCD", a grant from National Science Foundation of China (No. 11405104), the Ministry of Science and Technology, Taiwan under Grant No. 105-2112-M-002-017-MY3 and the Kenda Foundation. It stemmed from discussions during the ``Workshop of Recent Developments in QCD and Quantum Field Theories" at National Taiwan University in November, 2017. We would like to thank K. Orginos for useful comments and V. Braun and A. Vladimirov for useful discussions. 

\vspace{2em}

\begin{center}
\bf{Note added}
\end{center}

While this work is being submitted to arXiv, a preprint by A. Radyushkin~\cite{Radyushkin:2018cvn} appears, where a similar formula as our Eq.~(\ref{extraction}) has been presented, which in the latest version agrees with our result.

\end{document}